%% file: 2020_faraday_schroederturk.tex
\documentclass[twoside,twocolumn,9pt]{article}
\usepackage{extsizes}
\usepackage[super,sort&compress,comma]{natbib} 
\usepackage[version=3]{mhchem}
\usepackage[left=1.5cm, right=1.5cm, top=1.785cm, bottom=2.0cm]{geometry}
\usepackage{balance}
\usepackage{times,mathptmx}
\usepackage{sectsty}
\usepackage{graphicx} 
\usepackage{lastpage}
\usepackage[format=plain,justification=justified,singlelinecheck=false,font={stretch=1.125,small,sf},labelfont=bf,labelsep=space]{caption}
\usepackage{float}
\usepackage{fancyhdr}
\usepackage{fnpos}
\usepackage[english]{babel}
\addto{\captionsenglish}{%
  
}
\usepackage{array}
\usepackage{droidsans}
\usepackage{charter}
\usepackage[T1]{fontenc}
\usepackage[usenames,dvipsnames]{xcolor}
\usepackage{setspace}
\usepackage[compact]{titlesec}
\usepackage{hyperref}
\usepackage{lineno}

\usepackage{epstopdf}

\definecolor{cream}{RGB}{222,217,201}

\begin{document}

\pagestyle{fancy}
\thispagestyle{plain}
\fancypagestyle{plain}{

\fancyhead[C]{\includegraphics[width=18.5cm]{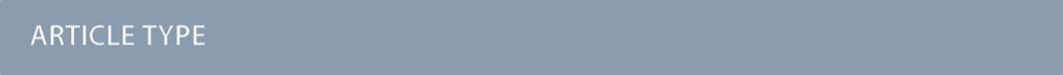}}
\fancyhead[L]{\hspace{0cm}\vspace{1.5cm}\includegraphics[height=30pt]{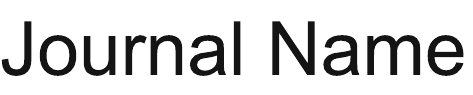}}
\fancyhead[R]{\hspace{0cm}\vspace{1.7cm}\includegraphics[height=55pt]{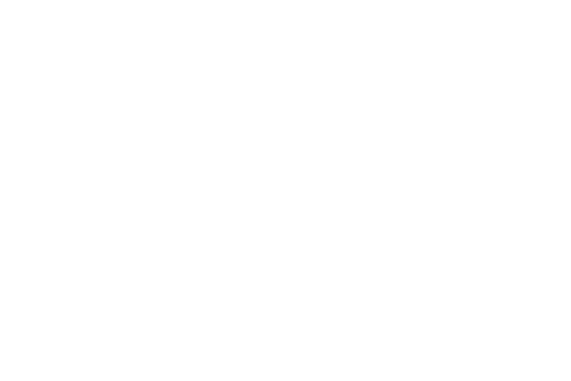}}
\renewcommand{\headrulewidth}{0pt}
}

\makeFNbottom
\makeatletter
\renewcommand\LARGE{\@setfontsize\LARGE{15pt}{17}}
\renewcommand\Large{\@setfontsize\Large{12pt}{14}}
\renewcommand\large{\@setfontsize\large{10pt}{12}}
\renewcommand\footnotesize{\@setfontsize\footnotesize{7pt}{10}}
\makeatother

\renewcommand{\thefootnote}{\fnsymbol{footnote}}
\renewcommand\footnoterule{\vspace*{1pt}%
\color{cream}\hrule width 3.5in height 0.4pt \color{black}\vspace*{5pt}} 
\setcounter{secnumdepth}{5}

\makeatletter 
\renewcommand\@biblabel[1]{#1}            
\renewcommand\@makefntext[1]%
{\noindent\makebox[0pt][r]{\@thefnmark\,}#1}
\makeatother 
\renewcommand{\figurename}{\small{Fig.}~}
\sectionfont{\sffamily\Large}
\subsectionfont{\normalsize}
\subsubsectionfont{\bf}
\setstretch{1.125} 
\setlength{\skip\footins}{0.8cm}
\setlength{\footnotesep}{0.25cm}
\setlength{\jot}{10pt}
\titlespacing*{\section}{0pt}{4pt}{4pt}
\titlespacing*{\subsection}{0pt}{15pt}{1pt}

\fancyfoot{}
\fancyfoot[LO,RE]{\vspace{-7.1pt}\includegraphics[height=9pt]{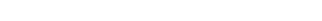}}
\fancyfoot[CO]{\vspace{-7.1pt}\hspace{13.2cm}\includegraphics{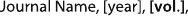}}
\fancyfoot[CE]{\vspace{-7.2pt}\hspace{-14.2cm}\includegraphics{RF}}
\fancyfoot[RO]{\footnotesize{\sffamily{1--\pageref{LastPage} ~\textbar  \hspace{2pt}\thepage}}}
\fancyfoot[LE]{\footnotesize{\sffamily{\thepage~\textbar\hspace{3.45cm} 1--\pageref{LastPage}}}}
\fancyhead{}
\renewcommand{\headrulewidth}{0pt} 
\renewcommand{\footrulewidth}{0pt}
\setlength{\arrayrulewidth}{1pt}
\setlength{\columnsep}{6.5mm}
\setlength\bibsep{1pt}


\makeatletter 
\newlength{\figrulesep} 
\setlength{\figrulesep}{0.5\textfloatsep} 

\newcommand{\topfigrule}{\vspace*{-1pt}%
\noindent{\color{cream}\rule[-\figrulesep]{\columnwidth}{1.5pt}} }

\newcommand{\botfigrule}{\vspace*{-2pt}%
\noindent{\color{cream}\rule[\figrulesep]{\columnwidth}{1.5pt}} }

\newcommand{\dblfigrule}{\vspace*{-1pt}%
\noindent{\color{cream}\rule[-\figrulesep]{\textwidth}{1.5pt}} }

\makeatother

\twocolumn[
  \begin{@twocolumnfalse}
\vspace{3cm}
\sffamily
\begin{tabular}{m{4.5cm} p{13.5cm} }

\includegraphics{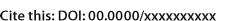} & \noindent\LARGE{\textbf{Quo vadis Biophotonics? Wearing serendipity \& slow science as a badge of pride, and embracing biology.} \linebreak {\small {Final version published as Faraday Discuss., 2020,223, 307-323, \href{https://doi.org/10.1039/D0FD00108B}{https://doi.org/10.1039/D0FD00108B}}}} \\
\vspace{0.3cm} & \vspace{0.3cm} \\

 & \noindent\large{Gerd E.\ Schr\"oder-Turk,$^{\ast}$\textit{$^{a}$}} \\

\includegraphics{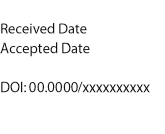} & \noindent\normalsize{This article is a reflection on the themes of the Faraday Discussion Meeting on 'Biological and Bioinspired optics' held from 20-22 July 2020. It is a personal perspective on the nature of this field as a broad and interdisciplinary field that has led to a sound understanding of the materials properties of biological nanostructured and optical materials. The article describes how the nature of the field and the themes of the conference are reflected in particular in work on the 3D bicontinuous biophotonic nanostructures known as the single Gyroid and in bicontinuous structures more broadly. Such single gyroid materials are found for example in the butterfly {\it Thecla opisena}, where the questions of biophotonic response, of bio-inspired optics, of the relationship between structure and function, and of the relationship between natural and synthetic realisations are closely interlinked. This multitude of facets of research of single gyroid structures reflects a beauty of the broader field of biophotonics, namely as a field that lives from embracing the serendipitous discovery of biophotonic marvels that nature offers to us as seeds for in-depth analysis and understanding. The meandering nature of its discoveries, and the need to accept the slowness that comes from exploration of intellectually new or foreign territory, means that the field shares some traits with biological evolution itself. Looking into the future, I consider that a closer engagement with living tissue and with biological questions of function and formation, rather than with the materials science of biological materials, will help ensure the continuing great success of this field. }

\end{tabular}

 \end{@twocolumnfalse} \vspace{0.6cm}

  ]

\renewcommand*\rmdefault{bch}\normalfont\upshape
\rmfamily
\section*{}
\vspace{-1cm}


\footnotetext{\textit{$^{a}$~Murdoch University, College of Science, Health, Engineering \& Education, 90 South St, Murdoch WA 6150, Australia. Tel: +61 8 9360 6350; E-mail: g.schroeder-turk@murdoch.edu.au}}





\section{Introduction}

The Faraday Discussion meeting 'Biological and Bioinspired optics', held virtually on 20-22 July 2020 and jointly with the 'Living Light' conference, provided an excellent discussion forum of research into biophotonics and bio-inspired optics, despite the organisational difficulties imposed by the COVID-19 pandemic.

Through its topics, the meeting was interdisciplinary in its nature, encompassing aspects of physics, materials science, chemistry, materials engineering and biology. Both by topic and by attendees, there was overlap with many recent conferences aimed at deciphering nanostructural optical mechanisms in biology, and butterfly coloration effects and butterfly(-inspired) nanostructures in particular.

\begin{figure*}[t]
  \includegraphics[width=\textwidth]{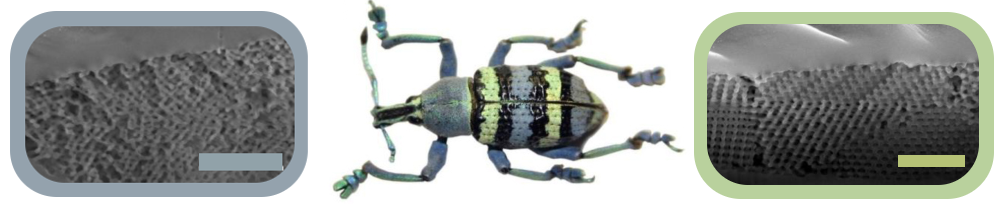}
  \caption{\label{pouya-beetle}The {\it Eupholus magnificus} beetle, as described by Pouya {\it et al.}\cite{Pouya:11} remains a magnificent example of photonic materials with varying degrees of order. The yellow part of the cuticle has an ordered Diamond nanostructure, whereas the blue part only shows quasi-order. Scale bars are 2 microns. (Images reproduced in adapted form from Pouya {\it et al.}\cite{Pouya:11})}
\end{figure*}

The meeting's first lecture, by Peter Vukusic, opened the discussion on the current focus of the field of biophotonics by reference to classical works  and by reference to the coherent core theme of 'structure versus function'.

The classical works Peter Vukusic referenced were the article by Lord Rayleigh '{\it On the optical character of some brilliant animal colours}'\cite{doi:10.1080/14786440108635867} and the book by Robert Hooke '{\it Micrographia, or some Physiological Descriptions of minute bodies made by magnifying glasses, with observations and inquiries thereupon}'\cite{hooke}. These are works by two scientists who, while polymaths, are probably most accurately characterised as physicists. Their interest was not so much in the biology as in using the biological structure (in this instance, thin layers) for the description of a physical phenomenom. 

Peter Vukusic also referenced the work of biologist Helen Ghiradella as inspirational and influential for the field, noting specifically her work 'Light and color on the wing: structural colors in butterflies and moths'\cite{Ghiradella:91}. Her work was a comprehensive study of butterfly nanostructure development using the tools of physics (i.e.\ electron microscopy), and has inspired related studies decades later.\cite{SaranathanOsujiMochrieNohNarayananSandyDufresnePrum:2010,Winter12911,Wiltse1603119}

Peter Vukusic then identified the question of structure versus function (or physical property) as one of the core themes of the field of biophotonics and bio-inspired optics. This, surely, has been an amazing success which is well evidenced by the understanding that has resulted from countless publications over the last two decades that have described the optical properties (the function) that results from a plethora of natural nanostructures including, as ordered examples, the {\it Morpho} butterfly christmas tree structures, the multi-layer reflectors, the Bouligand or chiral woodstack structures (see recent reviews\cite{SHARMA2014161,WILTS2014177}), opal-like or inverse-opal-like packings of sphere-like objects or cavities, hexagonal cylinder packings reminiscent of photonic fibres, three-dimensional ordered gyroid or diamond network structures, and others. More often than not, the optical effects were effects studied in dead tissues, such as feathers, insect cuticles, butterfly wing scales, compound eye structures, crustacean components or fish scales. For reviews see here\cite{mohan99,vukusic2003,doi:10.1002/lpor.200900018,doi:10.1002/adma.201705322}

Peter Vukusic placed particular emphasis on one aspect of the relationship between structure and function, namely the question of how relevant order or disorder in the structure is for the optical function of the nanostructure. Much of that discussion is captured in the article by S\'ebastien Mouchet {\it et al}, `{Optical Costs and Benefits of Disorder in Biological Photonic Crystals}'.\cite{D0FD00101E} It discusses the important question how robust the optical function of a nanostructure is to degrees of distortion or structural disorder; Biological nanostructures with their inherent imperfections continue to be important for this question. The beetle {\it Eupholus magnificus} remains a beautiful example of the possibility of photonic structures of varying degree of order, see Fig.\ \ref{pouya-beetle}. Aside from structures with imperfections, also strongly disordered structures have been shown to generate physical signals, particularly through scattering (such as in the brilliantly white beetles\cite{Vukusic348,doi:10.1002/adma.201702057}) or through omnidirectional bandgaps in hyperuniform structures \cite{Florescu20658}. 

Much (but not all\cite{Cuthilleaan0221}) of the research into structure-function relationships in biological nanostructures has been carried out from a perspective of materials science and physics and using the tools of those disciplines. This appears correct both for the research presented at the Faraday meeting, and also for other research in the biophotonics research community (for example going by research summarised in these special issues or conference proceedings\cite{doi:10.1098/rsif.2009.0013.focus,hyde2012,SARRAZIN2014107,doi:10.1098/rsfs.2017.0035}). The role of biology is, in these studies, mostly the provision of an 'encyclopedia' of amazing nanostructures (aside from nature's awe-inspiring and often mysterious mechanisms of producing such complex structures). 

\section{Scientific contributions at the meeting}

Over three days, the Faraday Discussion meeting continued with the discussion of 16 peer-reviewed manuscripts that were discussed in four sessions and with an additional poster session.

\begin{figure*}[h]
\centering
  \includegraphics[width=\textwidth]{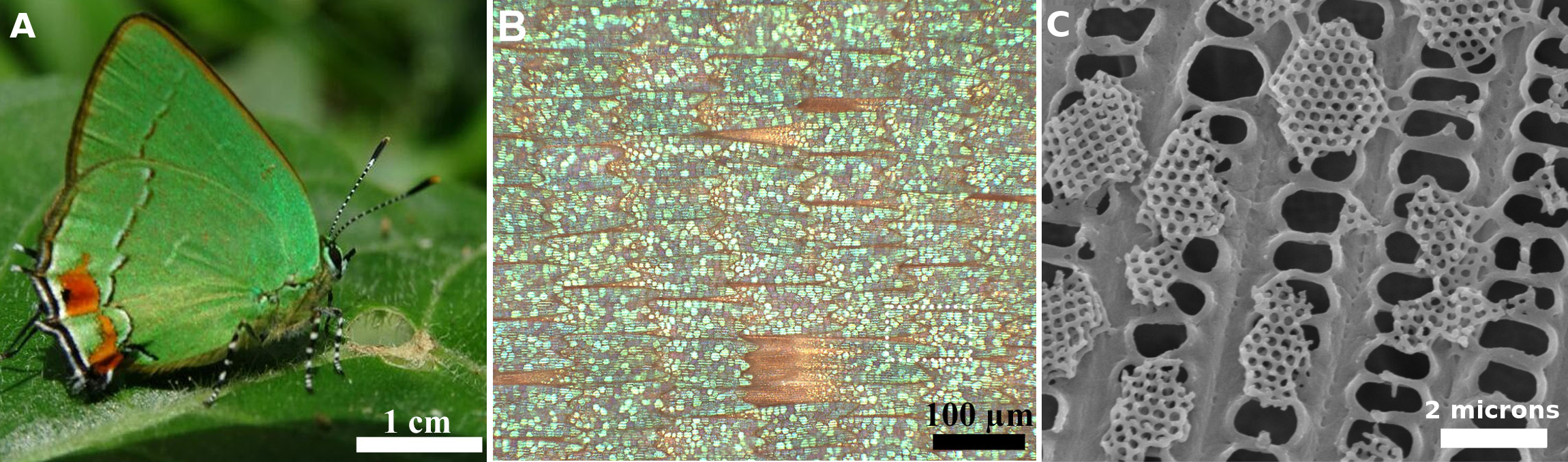}
  \caption{{\bf Optics \& Photonics in Nature}: The green speckled appearance of the wing-scales of {\it Thecla opisena} is due to crystallites with an internal three-dimensional nanostructure that can be modelled by the single gyroid geometry. (Data and images from the same samples and specimen as used in \cite{Wiltse1603119}.)}
  \label{fig:optics-photonics-in-nature}
\end{figure*}

In the theme {\it Optics \& Photonics in Nature}, physical analyses of structural aspects and optics of (dead) peacock feather nanostructures\cite{D0FD00033G}, scattering properties for UV radiation of (dead) egg shells\cite{D0FD00034E} and of epidermal cells in {\it Mandevilla} flowers\cite{D0FD00055H} were presented, as was a physical method (optical retinography) applied to the question of light sensing in insect eyes\cite{D0FD00075B}.

In the theme {\it Bioinspired Optics}, a change in photonic or plasmonic properties that results from the interaction of a bacterium with silver was used to construct a sensitive sensor for this bacterium\cite{D0FD00026D}. Wilts {\it et al.}\cite{D0FD00027B} used a trick, long in use in the wings of the cabbage butterfly to create high-dielectric constrast scatterers\cite{doi:10.1002/adom.201600879}, to create high-dielectric fluids; it makes use of the Kramers-Kronig relationship whereby UV absorbing pigments lead to high refractive index in the visible spectrum\cite{D0FD00027B}. A further article investigated optical gas sensing applications, with machine learning optimisation, using structures found also in butterflies but with little other relationship to biology\cite{D0FD00035C}. Finally, Franziska Schenk --an artist exploring the use of pigments based on structural color-- presented her beautiful work.\cite{D0FD00036A}. 

In the theme {\it The role of structure: order vs.\ disorder}, K\"ochling {\it et al.} presented a scaling analysis of height and spacing in the ridge structures in butterfly wing-scales and argue its relevance for non-coloration properties of the scale {\it bauplan} and functionality\cite{D0FD00038H}. Van de Kerkhof {\it et al.} showed that, in {\it Tradenscantia} leaves, a disordered layer of wax platelets as an epicuticular nanostructure provide the plant with a strong reflectance in the higher wavelength regime and is responsible for a golden shine.\cite{D0FD00024H} Martella {\it et al} described a liquid-crystal-based light-modulated 'actuating' mechanism for 'bio-synthetic' systems, such as for example nanostructures in butterflies.\cite{D0FD00032A}

Finally, in the theme {Natural and synthetic materials}, Chagas {\it et al.}\ described a 'coffee ring' mechanism for tuning structural coloration effects in cellulose-based 'cholesteric' arrangements.\cite{D0FD00020E} Goessling {\it et al.}\ discussed opportunities and limitations of the use of diatoms, or rather their biosilica shells, as alternatives to clean-room nanofabricated photonic crystals.\cite{D0FD00031K} Schiffmann {\it et al.}\ investigated a platelet reflector structure in the eye of a particular shrimp and concluded that its likely optical function was a camouflage effect to conceal the dark eye pigments in an otherwise largely transparent animal.\cite{D0FD00044B}  

Most of the work presented was in its nature materials science, materials chemistry or soft matter physics. The link to biology was, for most articles, either the analysis of a nanostructural phenomenom in a biological system or the inspiration from a previously described biological nanostructure or phenomenom. None of the studies addressed in any substantive way a biological structure-function relationship, that is, a link between the structure and its biological function or evolutionary benefit or cost (although some, such as\cite{D0FD00101E,D0FD00044B,D0FD00038H,D0FD00024H}, certainly provided information that would likely be very useful and integral for such a biological study). Few of the presenting authors were biologist. (The authors of \cite{D0FD00044B} are an exception, from a Structural Biology group, and the lead author of \cite{D0FD00075B} is a biologist by training.)

The above is not a valued statement; it is merely an observation about the content of a meeting characterised by outstanding science, and perhaps about the nature of the broader field of biophotonics and its community. A field that has links to biology, through the nanostructures and phenomena it addresses, but that is probably more accurately characterised by 'materials science applied to or inspired by biology' rather than 'biology done using materials science methods'. 

\section{Single gyroid biophotonic materials, and other bicontinuous structures}

The biophotonic crystals based on the single gyroid nanostructure are my personal area of biophotonic research and are an interesting example of the interwoven nature of the four threats of the field, reflected in the four sessions of the conference.

I will indulge a little in describing the research, including my own, related to the single gyroid nanostructure, its occurrence, formation and function in butterflies and its inspiration for optical and materials science research. We have come a long way in understanding many aspects of chemical synthesis, nanoengineering realisations, structure-function properties and optical properties of single gyroids (and related materials). Yet, we have thus far failed to obtain an equivalent understanding of the biological questions in relation to its formation mechanism and biological (or evolutionary) purpose, and have indeed struggled to even dedicate substantial collective effort to this question. This, I would argue, is despite the fact that materials science, membrane chemistry and soft matter physics are likely to be crucial in deciphering these questions and are likely to gain substantial further insight for bioinspired processes or designs.

In the field of biophotonics, the single Gyroid is by now well established as the nanostructural geometry of several green butterflies. In these butterflies, the single gyroid is a network-like, highly-ordered and chiral (spacegroup $I4_132$) geometry, which acts as a photonic crystal that produces a green coloration.\cite{doi:10.1098/rsif.2007.1065,SaranathanOsujiMochrieNohNarayananSandyDufresnePrum:2010,SCHRODERTURK2011290,C1CC11637A,doi:10.1098/rsif.2013.1029,Wilts681}

More broadly, the Gyroid (both in its form as the {\it $I4_132$ single gyroid} and as the {\it $Ia\overline{3}d$ double gyroid}) is a member --and perhaps the epitomy-- of a broader class of nanostructural geometries, known as bicontinuous geometries, see here for some reviews\cite{doi:10.1002/adma.201900818,HydeOKeeffeProserpio2008,1997iv,LUZZATI1997661}. These bicontinuous geometries are by now firmly established as a commonly observed structural motif in synthetic soft and hard nanostructured systems: Bicontinuous phases occur generically in synthetic self-assembled amphiphilic systems (lipids \cite{hyde1984,doi:10.1021/la5005837,doi:10.1021/la303833s} and copolymers \cite{Khandpur:1995}), in inorganic mesoporous solids \cite{AttardGlydeGoeltner:1995,AlfredssonAnderson:1996}, and  as 'cubosomes' in pharmaceutical drug-delivery systems \cite{SMLL:SMLL201300348,Drummond1999449}. Fifty years or so after the first reports of gyroid-like or bicontinuous phases in lipids (Luzzati's work on divalent cation soaps \cite{luzzati1967}), we have a pretty solid understanding of the formation and self-assembly mechanisms of these phases and of potential applications.  

In biological systems, bicontinuous phases, including gyroid-like geometries, have been observed ubiquitously in convoluted membranes in cell organelles\cite{Almsherqi2009275} (both plant cells and animal cells), biopolymeric nanostructures in butterflies and beetles \cite{nanoPrum2015}, fibril structures in keratin \cite{PhysRevLett.112.038102}, and lipid phases formed in the lung's alveolar surface and in human digestion models \cite{doi:10.1021/nn405123j}. However, in contrast to the synthetic system, I would argue that we have a far poorer understanding of the biological function and formation of such bicontinuous phases, even though the first description of these phases in biology is similarly old (Gunning's discussion of the structure of the prolamellar body in plant chloroplast precursors\cite{Gunning1965}).

\begin{figure*}[h]
\centering
  \includegraphics[width=\textwidth]{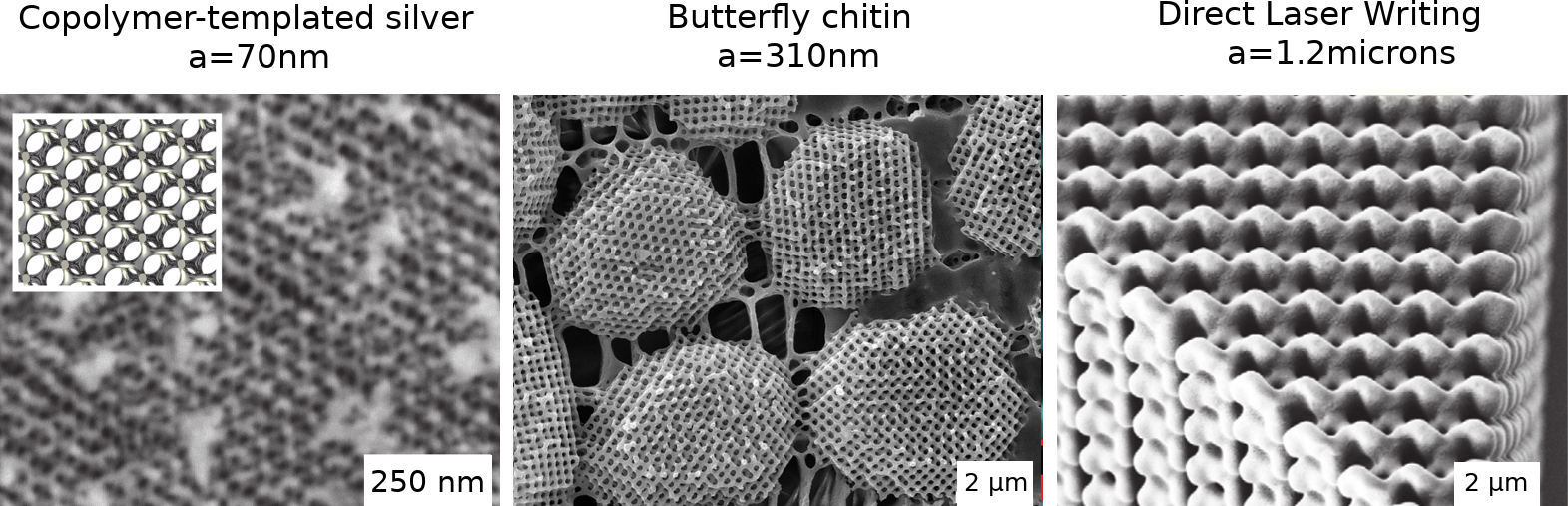}
  \caption{{\bf Natural and synthetic materials}: The {\it single gyroid} structure has been generated both as a synthetic material and as a natural material, using a variety of different processes and covering length scales from $nm$ to $cm$. (Left) Copolymer-templated silver nanostructure\cite{wiltscircularsilver2020}; (Middle) Chitin single Gyroid in the green butterfly {\it Thecla opisena}\cite{Wiltse1603119}; and (right) nanofabricated single gyroid as an optical device prototype \cite{TurnerSabaZhangCummingSchroederTurkGu:2013}. (Images reproduced from publications \cite{wiltscircularsilver2020,Wiltse1603119,TurnerSabaZhangCummingSchroederTurkGu:2013} with permission of the authors)}
  \label{fig:synthetic-vs-natural}
\end{figure*}

In terms of the optical properties, the last decade or so has seen the emergence of a small field of 'gyroid photonics', that is, the optics of materials with relevant optical, photonic or plasmonic properties related to a gyroid-like or gyroid-inspired structure. Particular interest in these geometries was driven by experimental realisations in the right length-scale regime for optical properties. Aside from the natural realisations in butterflies, these structures result from the ability to fabricate or synthetize Gyroids in a variety of systems, over a whole range of length scales, and including both top-down and bottom-up processes (see also figure \ref{fig:synthetic-vs-natural}): At the scale of millimeter and centimeter, three-dimensional printing and additive manufacturing has been used to study photonic crystals in the microwave range\cite{pouyasquash,doi:10.1098/rsfs.2011.0091,Hielscher:17}. Nanofabrication and direct laser-writing has enabled the creation of single gyroids with length scale smaller than the butterfly \cite{Gane1600084} or at slightly larger length scales corresponding to the telecom-band \cite{TurnerSchroederTurkGu:2011,TurnerSabaZhangCummingSchroederTurkGu:2013}. Nanofabrication has been used to produce single gyroids in both polymeric materials\cite{TurnerSchroederTurkGu:2011,TurnerSabaZhangCummingSchroederTurkGu:2013}, high dielectric-contrast glasses\cite{Cumming:14} and, after coating, in silver\cite{Cumming:18}. Gyroids with the same symmetry as the single gyroid can also be self-assembled in linear ABC terpolymers\cite{doi:10.1021/ma0493426}, at length scales $a\approx 70 nm$, and these self-assembled structures can be used to create gyroid-structured thin film materials in both gold\cite{doi:10.1002/adma.201103610} and silver\cite{wiltscircularsilver2020} with interesting optical properties, see also recent reviews on these matters\cite{ADOM:ADOM201400333,doi:10.1002/adma.201705708}.  This field has shed substantial light on the gyroid and its relatives as an optical material, both experimentally\cite{weyl,TurnerSabaZhangCummingSchroederTurkGu:2013,doi:10.1002/adma.201803478,wiltscircularsilver2020} and theoretically\cite{PhysRevB.88.245116,cryst5010014,SabaThielTurnerHydeGuBrauckmannNeshevMeckeSchroederTurk:2011,PhysRevApplied.2.044002,doi:10.1021/acsphotonics.6b00400,doi:10.1002/adma.201202788}, and has established the gyroid (particularly the single gyroid) as a somewhat widely known photonic design (see e.g.\ the review by Dolan {\it et al}\cite{doi:10.1002/adom.201400333}).

What about the biological relevance of the butterfly gyroids? What do we know about the relevance of the gyroid for the evolution of the butterflies? The honest answer is 'not very much'. We understand that the circular-polarisation effects due to its chirality are unlikely to play a significant role\cite{Saba2014193} (note also the work by Corkery {\it et al}\cite{doi:10.1098/rsfs.2016.0154} with a potentially different conclusion). We also understand the interplay of pigmentary and structural effects\cite{Wilts681}. And there is a strong correlation between structural and pigmentary traits (optical properties) and the taxonomical distribution (at least for the {\it Parides} species\cite{Wilts2014taxo}). However, it appears that we have little understanding of e.g.\ the difference between a multi-layer reflector and a gyroid nanostructure in terms of evolutionary benefit or cost. 

The single gyroid exemplifies the materials science approach that is perhaps prevalent in the biophotonics community: We have embraced biophotonic nanostructures and mechanisms wholeheartedly as a materials science problem, and have excelled in developing an understanding of all materials science aspects of these systems: from optical properties, to mechanical structure-property relationships\cite{kapfer2011b}, to self-assembly mechanisms and the use of natural designs for nanoengineering applications. Using natural nanostructures to conduct and to inspire amazing materials science research is the true success story of the field, and well reflected in the themes of this 'Biophotonics and bio-inspired optics' meeting. 

Yet, we have not managed to connect deeply to the biological questions, such as with regards to formation mechanisms or with regards to evolutionary biological structure-function relationships.

\begin{figure}[th!]
\centering
  \includegraphics[width=\columnwidth]{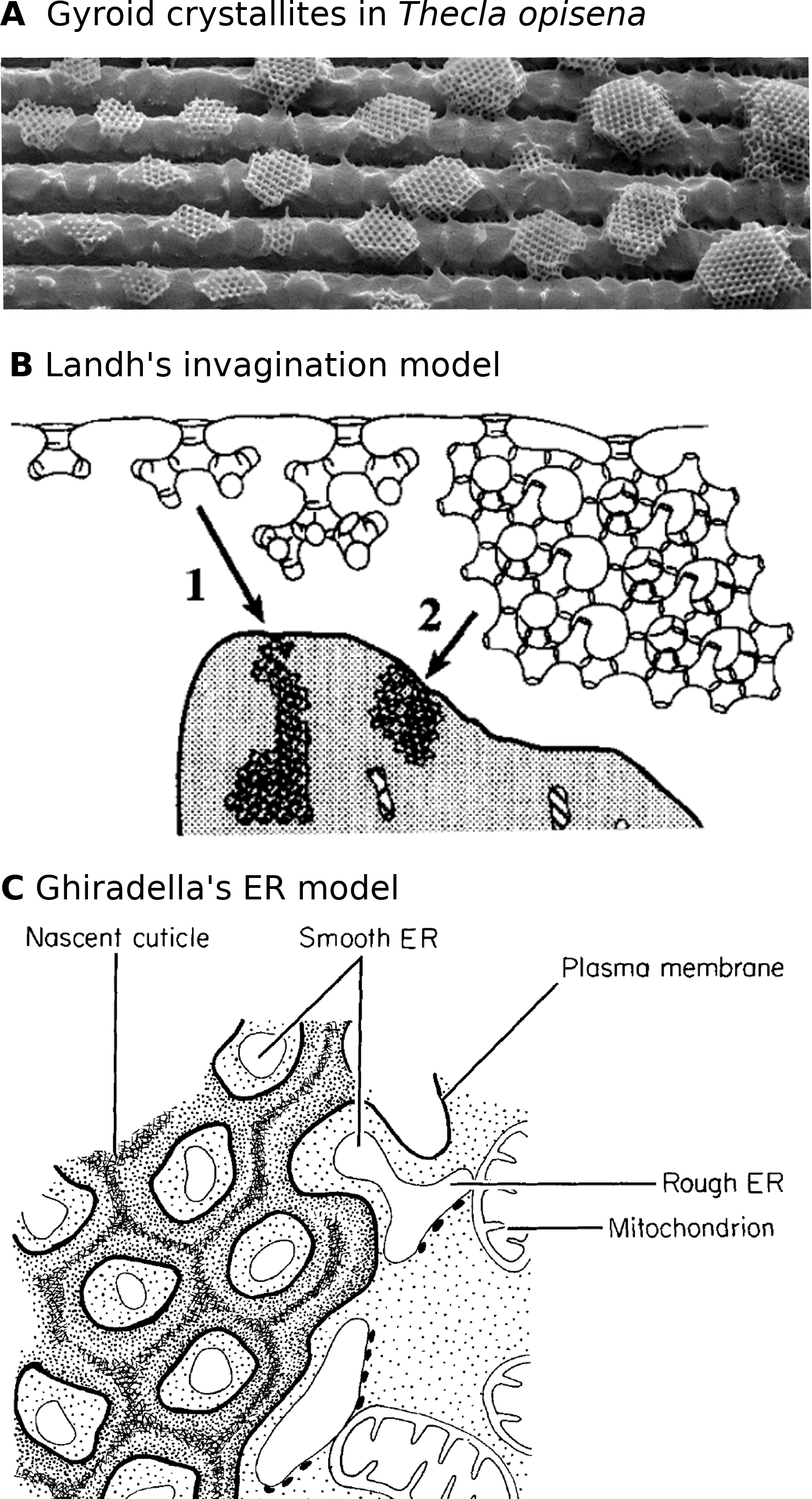}
  \caption{\label{fig:landh-vs-wilts}{\bf Towards {\it in-vivo} imaging of butterfly nanostructural development}: Understanding the growth mechanism of single gyroids in green butterflies will be likely achieved through a combination of advances in structural and developmental cell biology, biochemistry, materials science and nanostructural microscopy. We will have to see how the gyroid crystallites found by Wilts {\it et al.}\cite{Wiltse1603119} will relate to Helen Ghiradella's model based on membranes in the endoplasmic reticulum\cite{doi:10.1002/jmor.1052020106} and to Tomas Landh's invagination model\cite{doi:10.1016/0014-5793(95)00660-2}. Images reproduced in adapted form from Wilts {\it et al}\cite{Wiltse1603119} (top), Landh\cite{doi:10.1016/0014-5793(95)00660-2} (middle) and Ghiradella\cite{doi:10.1002/jmor.1052020106} (bottom).}
  
\end{figure}

Specifically for the single gyroid nanostructure in butterflies, we do not fully understand the mechanisms by which this structure emerges. There has been the seminal work by Ghiradella\cite{Ghiradella:91,doi:10.1002/jmor.1052020106}, referenced in the opening lecture, which used electron microscopy to assert that a folding of the plasma membrane into a nanostructured form is involved. There have been similar more recent studies.\cite{SaranathanOsujiMochrieNohNarayananSandyDufresnePrum:2010,Winter12911,Wiltse1603119} These include our recent suggestion to infer aspects of the formation mechanism from the fortuitous finding of a blob-like gyroid structure in {\it Thecla opisena}, see figure \ref{fig:optics-photonics-in-nature} and figure \ref{fig:landh-vs-wilts}. Nevertheless, a full understanding of this growth process and its relationship\cite{wilts_clode_patel_schröder-turk_2019} to the self-assembly processes of synthetic gyroids will, in my opinion, only develop in closer collaboration with biology and will benefit from the advances in imaging (e.g.\ evident in the recent work on F-actin's role in wing-scale development\cite{DINWIDDIE2014404}).   

Why should a researcher with a materials science background take an interest in the biological development of the gyroid nanostructure in butterflies? The answer to this question is at least two-fold:\ First, because we can! The elucidation of the growth process of this structure in butterflies is likely to benefit from the concepts, techniques and measurements that are inherent to soft matter physics, materials chemistry and materials science. Specifically for the gyroid development in butterflies, one hypothesized development route is by a templating membrane (part of the endoplasmic reticulum) with chitin expression and polymerisation inside the mould provided by that template. This has some similarities to the crystal growth of a calcite crystal that takes place inside a copolymeric gyroid template\cite{doi:10.1002/adma.200900615}. The question of the interplay of self-assembly transitions and growth processes\cite{wilts_clode_patel_schröder-turk_2019} in the biological system is one that materials scientists are well equipped to address.

A second argument is that there are many insights to be gained from nature in relation to assembly mechanisms for gyroid nanostructures. Nature, in the form of butterflies or through intracellular membranes, achieves a number of feats that self-assembly strategies have as yet failed to achieve (despite recent progress in stabilising larger lengthscale gyroid\cite{PMID:25790335}). These include the very large length scales of the nanostructures and cubic membranes found in biology (e.g.\ circa 300 nm for the butterfly single gyroids and 400 nm for some gyroid-structured membranes in the mitochondria in retinal cone cell of the rodent {\it Tupaia belangeri}\cite{Almsherqi2009275}); the ability to for single network structures or unbalanced membranes with one channel of greater volume fraction than the other (e.g.\ in prolamellar bodies in plants\cite{doi:10.1002/adma.200900615}); the ability for enantiomer-specific chiral selection (such as the preference for one of the two chiralities of the single gyroid in butterflies\cite{Winter20102015}).   

50 years of research into bicontinuous structures has been an incredible success story which has elucidated many of aspects of the amazing formation mechanisms and properties of these beautiful geometries. Much of that research has been achieved through interdisciplinary research spanning physics, mathematics, chemistry and engineering, and to a lesser extent biology; it appears befitting that the Gyroid's father and inventor, Alan Schoen, is a person who is not easily classified as a mathematician or physicist or engineer and with a strong appreciation of all these disciplines\cite{doi:10.1098/rsfs.2012.0023}. At least starting with the discovery of the gyroid in butterfly wing-scales by Michielsen\&Stavenga\cite{doi:10.1098/rsif.2007.1065}, the field of biophotonics has substantially contributed to this success of the gyroid story, adding a bulk of knowledge about its optical properties and uses.

I am led to believe that the next chapters of the gyroid story (and of other bicontinuous structures) will be written through a closer integration with biology.\cite{doi:10.1098/rsfs.2012.0035,doi:10.1098/rsfs.2017.0035} 

\section{Embracing serendipity and biology}

In the closing lecture (which forms the basis of this article) I quoted Doekele Stavenga, from an electronic correspondence he had sent me a day before, as saying

\begin{quote}
  It strikes me that the question of structure and function, a persistent question in biology, was frequently creeping up [in the Faraday meeting], although the audience or at least the presenters are mainly physicists/chemists and not biologists.
\end{quote}

and then further

\begin{quote}
  The answer to that question is only clear when the consequences for survival are known, something which is virtually always remote at this moment, because of the complexity and the numerous possible functions of one and the same structure. So, what we are left with is still (beautiful) stamp collection, i.e.\ studying special cases, or at most speculations about functions and the possibility to inspire applications.
\end{quote}

I consider these to be valid observations that the biophotonics research community should embrace, in parts as a matter of professional pride and in parts as guidance by one of the field's most senior scientist about a useful future direction. 

The observation that the community --as for example illustrated by the author list in this Faraday Discussion volume-- is predominantly composed of physicists, chemists and materials scientists appears accurate, and has been discussed above. 

I share the view that the biophotonics community and the field of biophotonics should, into the future, seek a closer connection to biology, in terms of the conceptual questions we address and in terms of the collaborations. To seek to engage with nature's nanostructures in living organisms, rather than in dead probes or specimen, is an imperative that I consider useful. It serves us well to remember that nature's solutions have often evolved over millions of years, resulting in clever, intricate, subtle and sustainable designs.\cite{Parker_2000}

Does that mean that biophotonic science without a strong focus on biological structure-function relations is stamp collecting or, pejoratively, 'just' stamp collecting?

When I referred to Stavenga's quote in my lecture, I did so with the intention of conveying two general sentiments: On the one hand, to express the above slight 'nudge' that a stronger engagement with biological evolutionary design is desirable and would give the field a coherent question into the future. Yet, on the other hand the positive connotation of stamp collecting as an exercise characterised by the serendipity inherent to collecting exercises, by an appreciation of the diversity of the collected items or discoveries, and by a willingness to engage deeply and slowly with details the importance of which a rushed observer would fail to understand. (I suppose, a loose description of all of this is 'nerdiness'.)

These positive features are certainly ones that I associate with a lot of science in general, and the field of biophotonics in particular. One could say that an oft applied approach of our field is to jump at the discovery of a natural optical structure or phenomenom --be this from serendipitous discovery or from systematic search-- and then get to the bottom of the physical basis and applications of the effect. This leads to a breadth and depth which I consider both enjoyable and fruitful for good solid science. The breadth and serendipity also leads to a breadth and versatility of the tool box, a further enjoyable and useful aspect of the field of biophotonics that sets it apart from other fields with a more developed but more monotonous, and sometimes even nearly industrial tool kit.

Hence, the loose association of the nature of our field with stamp collecting is one that recognises the positives of the slow and detailed nature of stamp collecting.

My suggestion to orient our mind-set closer to biology is not a criticism of past research of the field, but rather an informed opinion as to a desirable future direction for a mature field. 

How then do we embrace biology more? This is easier said then done, at least for the beautiful butterflies, moths, birds and beetles that have provided so much inspiration for the biophotonics field over the last few decades. While color variations can be achieved in intergenerational experiments, and hence through breeding, an experiment aiming to assess the true evolutionary cost or benefit of e.g.\ a nanostructural geometry (i.e.\ of a truly biological structure-function relationship) is difficult. This is especially true when sexual selection and hence the question of beauty, or rather perception of beauty, becomes relevant.\cite{PrumBeauty}

Perhaps, the understanding of vision and eye function in insects, crustaceans and other animals, is a great example of how biology and physics (or materials science) can jointly co-evolve to lead to substantive understanding of an entire research area. This achievement is due to the nature of this problem, with essential aspects that require both biology and advanced optics.\cite{STAVENGA2006307,Horvath,marshal2008,D0FD00075B,10.3389/fevo.2016.00018}

I consider that there are two core aspects to enhancing collaborative work between biology and physics (or materials science or materials chemistry): One is the broader issue how we enable slower and less immediately productive science, and that is covered in parts in the next section. The other, far more practical, is the choice of research question and research object (aka organism). A choice that lends itself more readily to both biological and materials science questions is likely to lead to greater successes.

One aspect that makes a system more amenable to both physics and biology, or that at least allow for more definite conclusions, is the degree of complexity. If the potential function of a structure can be reasonably assumed to relate to largely a single biological function, conclusions are easier to draw. This is for example the case for the heat-shielding and -emitting properties in Saharan silver ants\cite{Shi298}, or in the properties of crustaceans eyes \cite{D0FD00044B} or insect eyes\cite{BeluStavi2013}, or in (artificial) flower signals to bees\cite{bluehalo}.

The task of understanding the formation and function of nanostructure in living tissue  --presumably for all sciences-- is made simpler by having simple organisms or structures, easy accessibility to the section of the organism of interest and slow or continuous growth mechanisms. Such systems appear to include the collective structural coloration systems in bacteria colonies\cite{vukubact,ingham2020}; An easy and exciting (and perhaps already accomplished) experiment would be to see if a particular collective behaviour can be 'bred' into the colony through a tunable food supply with optical feedback that rewards a particular optical signal with greater or better food.

Similarly, when studying the genesis of structure in nature, i.e.\ the formation mechanisms and growth processes, a system amenable to both biological and physical analysis is useful. In general, nanostructure development in simple plants is more easily accessible than more complex organisms, both for relatively simple lamella stacks\cite{kirkdainius} and for bicontinuous geometries\cite{Kowalewska875}, or in other simple organisms such as ameoba\cite{dengamo}.

I believe there is significant value in making a concerted effort to identify, and then pursue, biological systems that lend themselves to and can benefit from a conjoint interdisciplinary analysis. A first, and slow, step in this is to immerse oneself in each other's communities.

\section{Slow science, empowered scientists}

I will finish this article --or opinion piece rather-- with some quotes from Saint Exupery's novel '{\it Le petit prince}'\cite{PetitPrince}. Perhaps these quotes just represent my deluded and romantic longing for a type of science that the university system I work in no longer favours; as such it could be a eulogy to a part of my scientific career which I have loved deeply and believed in firmly. Yet, I am led to believe that it is rather a sentiment shared by many others, and as such a (peaceful) call to arms to protect a system we believe serves our society well. 

In chapter 21 of the novel, the little prince encounters a fox who asks the petit prince to 'tame' (in French, {\it apprivoiser}) him. The little prince declines, with reference to his busy schedule and lack of time and to the multitude of things he has to discover: 

\begin{quote}
Je veux bien, r\'epondit le petit prince, mais je n'ai pas beaucoup de temps. J'ai des amis \`a d\'ecouvrir et beaucoup de choses \`a conna\^itre.
\end{quote}

The fox responds to say that one only truly knows the things (in French, {\it choses}) that one has tamed (In this context, the French word {\it choses} for things appears to include both living beings and objects):

\begin{quote}
  On ne conna\^it que les choses que l'on apprivoise, dit le renard.
\end{quote}

The meaning of the French word 'apprivoiser' is only poorly captured by 'tame', and its meaning is explored in the dialogue between the fox and the prince. It does not have the connotation of dominance that is inherent to the word {\it tame}, but rather embodies the deep engagement and a profound understanding and knowledge. It also embodies a sense of responsibility towards the tamed object and a sense of commitment to dedicate time and energy to the relationship. It certainly embodies a sense of befriending and the development of a tender and special relationship, and one that makes a particular object stand out from other objects of the same type. 

The fox continues to lament that mankind's lack of time has taken away the ability to get to know things in the way the word {\it apprivoiser} implies, that is, that create profound acquaintance and friendship. The fox laments the superficiality of human interactions, almost as a buy-and-dispose mentality:  

\begin{quote}
  Les hommes n'ont plus le temps de rien conna\^itre. Ils ach\`etent des choses toutes faites chez les marchands. Mais comme il n'existe point de marchands d'amis, les hommes n'ont plus d'amis. Si tu veux un ami, apprivoise-moi!
\end{quote}

The little prince realises that his relationship with a particular rose meets the criterion of {\it apprivoiser}. The fox describes how the time the prince has 'lost' on his rose is what makes it special
\begin{quote}
  C'est le temps que tu as perdu pour ta rose qui fait ta rose si importante.
\end{quote}

The dialogue ends with the prince recognising the emotional attachment to the fox that resulted from him taming the fox and that will result in pain when the prince leaves the fox:

\begin{quote}
  Mais tu vas pleurer!, dit le petit prince.\\ 
  Bien s\^ur, dit le renard.
\end{quote}

Does the relationship between the fox and the little prince bear any resemblence to how scientists relate to their objects of study? Or should they? 

In many aspects '{\it Surely not!}': Objectivity is a core pillar of any scientific discovery and emotional bias a hindrance towards any exploration of the truth. A nanostructure does not become more important than any other one just because a scientist has fallen in love with it. Just because an individual researcher or a community have 'wasted' a lot of time or resources on a topic, the topic does not become special and certainly does not deserve a further waste of time or resources. These are considerations that we, as researchers, need to keep an active awareness of. Indeed, we should scrutinise our work, and our community's work, with respect to these considerations. And, we should demand from the system we work in that it empowers us to do so!

\begin{figure}[t!]
\centering
  \includegraphics[width=\columnwidth]{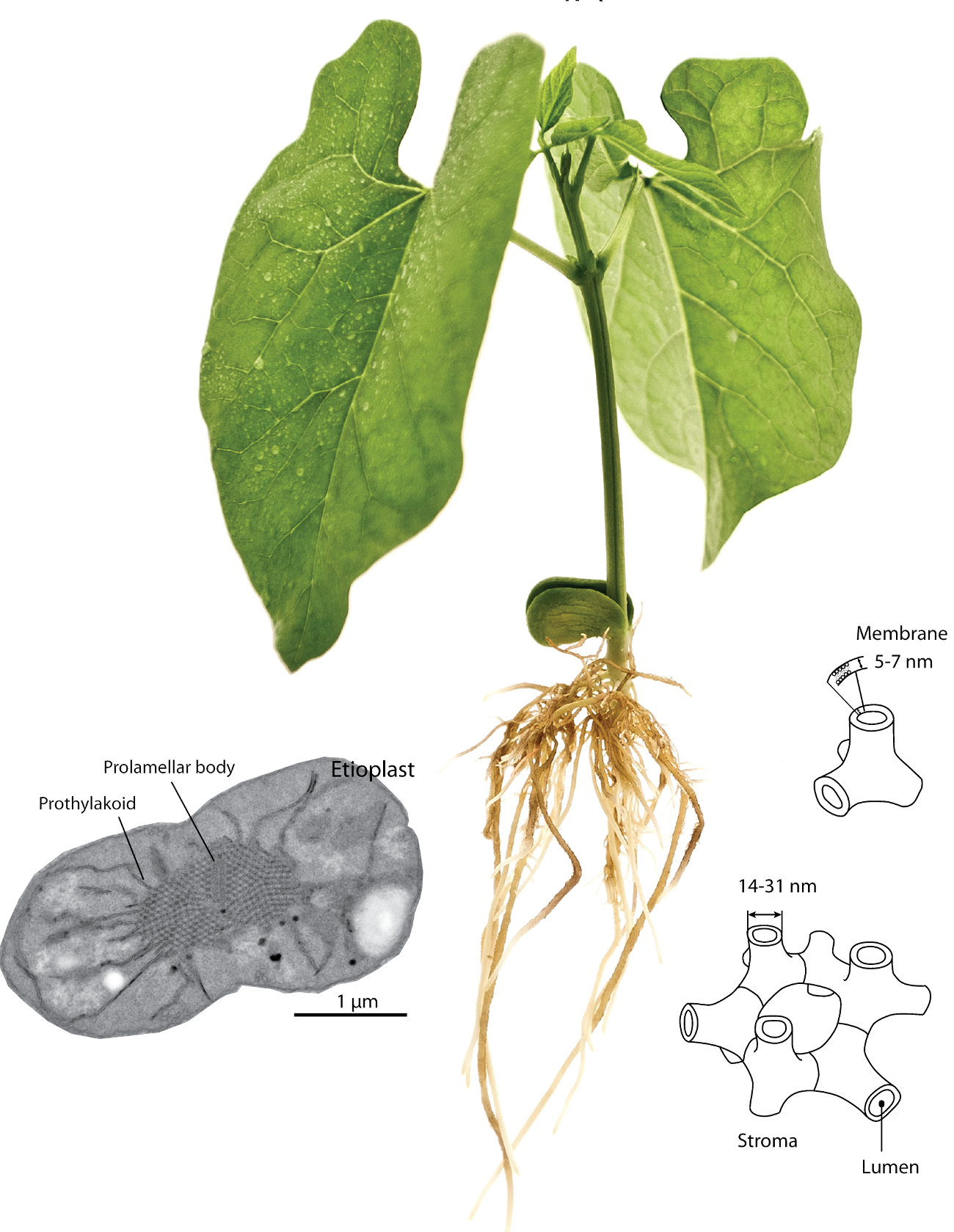}
  \caption{{\bf Is the next rose the runner bean?}: The runner bean {\it Phaseolus coccineus} is one of the many plants that, when grown in darkness, adopts a curious bicontinuous membrane structure in the chloroplast precursor, called the pro-lamellar body (curiously, often with the symmetry of the {\it single} diamond; that is, it is unbalanced). Nanostructural and light-related effects in plants, such as this and others, might lend themselves more easily to close collaborations between materials science, physics and biology, in particular given advances in imaging such materials\cite{Kowalewska875,doi:10.1080/23818107.2019.1619195,10.1093/jxb/erz496}. (Microscopy, photography and image composition by Lucja Kowalewska and Dainius Jakubauskas. Image adapted from Mezzenga {\it et al}.\cite{doi:10.1002/adma.201900818}) }
  \label{fig:runnerupbean}
\end{figure}

In other aspects '{\it Absolutely!}'. The profound and comprehensive understanding that we as scientists aim for requires a commitment and a dedication to a research object or topic, and to getting to know the ins and outs of the object, that has resemblances with the relationship embodied in the word {\it apprivoiser}. In fact, we identify poor science quickly when a study lacks depth and rigour, or when it stops with the proverbial low-hanging fruit rather than elucidate the subject with a breadth required for a full understanding. We recognise and appreciate the science that is informed by profound and comprehensive engagement with the topic, both in breadth and depth, and the scientists who produce such science (while, of course, recognising that occasional scientific flings are not contrary to a life of scientific virtue).

A key aspect of the above conflict is in the value we attach to these long-term friendships, the way we communicate about them and the way we 'show off' about the friendships and relationships. From the little prince's description of his relationship to the fox and the rose, one gets an acute sense of honesty, humility and openess. The little prince does neither talk up his relationship, nor does he talk it down. It does not appear that the little prince would measure the depth of his friendship to his rose by the number of people who applaud it, or indeed by the number of times he describes the friendship himself. I doubt that he would pay little attention to the superficial judgment of an outside as to the relative value of his tamed friendships.

Should a scientist love his or her science? Or is that the problem with us scientists? Perhaps the problem of the relationship between a scientist and his or her object of enquiry needs to be scrutinised against a famous claim in the Petit Prince novel that you can only see well with the heart:

\begin{quote}
Voici mon secret. Il est tr\'es simple: on ne voit bien qu'avec le c{\oe}ur. L'essentiel est invisible pour les yeux.
\end{quote}

So true, yet so wrong. Scientists are constantly confronted with the situation of needing to put their heart and soul into a research topic, yet to maintain an objective view of the object of their study and of its place in the broader scheme of things. In order to be the best scientists we can be, we need to maintain an acute awareness of this potential conflict.

Yet, we also need to develop the broad shoulders to demand an academic system that supports us scientists in managing this conflict. A system that enables our academic independence of our research topic yet that enables the deep engagement, the heart and soul, the love of a subject matter needed to achieve accurate, relevant and independent research outcomes of long-lasting value. This, ultimately, is not about a professional privilege but about the benefit of the society we serve.

In any case, the little prince's key message is that the ability to tame, the ability and willingness to get to know something or someone profoundly, is what leaves long-lasting impressions that stay forever. The strength of our research community --biophotonics or beyond-- comes from that commitment to profound research, to in-depth and long-term commitment and to communication of our findings that is honest. Lasting friendships, more so than fleeting love affairs. This has served the little prince well in relation to his rose, and it has served the biophotonics community well. We are good at 'taming' research topics, not flirting with them, and should wear this approach as a badge of professional honour and pride. 

Back to the prince's rose and the prince's fox. Which object should we tame? I predict that many of the future successes of the field of 'biophotonics and bio-inspired optics' will arise from us embarking on the slow journey of truly and deeply engaging with biology. Let's think carefully about which flower, pee, amoeba, bacterium, berry, butterfly or crustacean would be good to tame. 

\section*{Conflicts of interest}
There are no conflicts to declare.

\section*{Acknowledgements}

I wish to express my gratitude to Silvia Vignolini, Leila Deravi, Matthias Kolle, Lini Li and Bodo Wilts for organising an excellent Faraday Discussion Meeting '{\it Biophotonics and bio-inspired optics}'. 

I acknowledge support by the Australian Research Council, through grant DP200102593 {\it Meta-microscopy of insect tissue: How nature grows bicontinuous nanosolids}.

I am grateful to Bodo Wilts for many years of collaboration on Gyroid butterfly nanostructures, for joint work on the above ARC grant proposal from which some sections of this paper's text derive and for the electron microscopy image shown in Fig.\ \ref{fig:optics-photonics-in-nature}. I am grateful to Lucja Kowalewska and Dainius Jakubauskas for discussions of prolamellar bodies in plant plastids and for creating the illustration of the runner bean in Fig.\ \ref{fig:runnerupbean}.

I am grateful for the serendipity that led me into Stephen Hyde's research group in 2000, hence facilitating my introduction to the Gyroid. I am grateful to Stephen Hyde and Klaus Mecke for two decades of mentorship, support and friendship.

I am grateful for the serendipity that led Matthias Saba to join my research group in 2009 or so, and to Matthias Saba for embracing the topic of gyroid photonics with a depth and intellect that exceeded my own and that underpinned much of the success of our joint projects.

I wish to apologise that neither the presentation I gave at the meeting nor this paper have the high quality I expect of myself. Workplace-related stress has negatively impacted --hopefully temporarily-- my ability to produce high-quality research outputs.

I wish to apologise specifically to Peter Vukusic and Doekele Stavenga that I did not get a chance to show them drafts of my presentation prior to the presentation even though it contained substantial reference to their work and presentations they had given; The presentation and this article represent my own opinion, and I did not seek to imply that my opinion be shared by either of them. I apologise if the inclusion of their photos may have suggested otherwise. I would also like to apologise to the Faraday Discussion staff for very poor communication on my part and unacceptable delays that they nevertheless kindly accommodated. 

I thank Doekele Stavenga, Lucja Kowalewska and Bodo Wilts for comments on the manuscript.


\balance


\newpage

\onecolumn
\input{2020_faraday_schroederturk.bbl}

\end{document}

%% file: 2020_faraday_schroederturk.bbl
\providecommand*{\mcitethebibliography}{\thebibliography}
\csname @ifundefined\endcsname{endmcitethebibliography}
{\let\endmcitethebibliography\endthebibliography}{}